\documentclass[journal]{IEEEtran}
\usepackage{graphicx}% Include figure files
\usepackage{dcolumn}% Align table columns on decimal point
\usepackage{bm}% bold math
\usepackage[subfigure]{graphfig} %插多张图片的宏包
\usepackage{booktabs}            %插表格用的宏包
\usepackage{diagbox}             %插表格用的宏包
\usepackage{multirow}            %插多行表格用的宏包
\usepackage{pgfplots}
\pgfplotsset{width=9cm,compat=1.6}
\usepackage{import}
\usetikzlibrary{positioning}
\usepackage{float}

\begin{document}

%\preprint{APS/123-QED}

\title{Reducing the X-ray radiation exposure frequency in cardio-angiography via deep-learning based video interpolation }% Force line breaks with \\

\author{\IEEEauthorblockN{Xiao-Lei Yin\IEEEauthorrefmark{1},
		Dong-Xue Liang \IEEEauthorrefmark{1},
		Lu Wang\IEEEauthorrefmark{1}, 
		Jing Qiu\IEEEauthorrefmark{1},
		Zhi-Yun Yang\IEEEauthorrefmark{3}
		Jun-Hui Xing \IEEEauthorrefmark{4}
		Jian-Zeng Dong \IEEEauthorrefmark{3}
		 and Zhao-Yuan Ma \IEEEauthorrefmark{1}
		}
	\IEEEcompsocitemizethanks{
		%\IEEEcompsocthanksitem Manuscript received February 10, 2020.
		\IEEEcompsocthanksitem Dong-Xue Liang is with The Futrue Laboratory, Tsinghua University, Beijing, 10084, China.(E-mail: liang\_laurel@tsinghua.edu.cn)
		%\protect\\
	%	\IEEEcompsocthanksitem Xiao-Lei Yin and Lu Wang are with Academy of Arts and Design, Tsinghua University, Beijing, China.
		\IEEEcompsocthanksitem Xiao-Lei Yin, Lu Wang, Zhao-Yuan Ma and Jing Qiu are with  The Futrue Laboratory, Tsinghua University, Beijing,  China.
		\IEEEcompsocthanksitem Zhi-Yun Yang and Jian-Zeng Dong are with Center for Cardiology, Anzhen Hospital, Capital Medical University, Beijing, China.
		\IEEEcompsocthanksitem Jun-Hui Xing is with The First Affiliated Hospital of Zhengzhou University, Zhengzhou, China.
	}
}

%\markboth{ IEEE Transactions on Biomedical Engineering,~February 3, ~2020}%
%{ \MakeLowercase{\textit{et al.}}: }



\maketitle             %  but any date may be explicitly specified

\begin{abstract}

%本文的方法是在DAIN算的基础上，
Cardiac coronary angiography is a major technology to assist doctors during cardiac interventional surgeries. Under the exposure of X-ray radiation, doctors inject contrast agents through catheters to determine the position and status of coronary vessels in real time. To get a coronary angiography video with a high frame rate, the doctor needs to increase the exposure frequency and intensity of the X-ray. This will inevitably increase the X-ray harm to both patients and surgeons. In this work, we innovatively utilize a deep-learning based video interpolation algorithm to interpolate coronary angiography videos. Moreover, we establish a new coronary angiography image dataset ,which contains 95,039 triplets images to retrain the video interpolation network model. Using  the retrained network we synthesize high frame rate coronary angiography video from the low frame rate coronary angiography video. The average peak signal to noise ratio(PSNR) of those synthesized video frames reaches 34dB. Extensive experiment results demonstrate the feasibility of using the video frame interpolation algorithm to synthesize continuous and clear high frame rate coronary angiography video. With the help of this technology, doctors can significantly reduce exposure frequency and intensity of the X-ray during coronary angiography.

\end{abstract}
\IEEEpeerreviewmaketitle
      
%\tableofcontents

\section{Introduction}
Cardiac coronary angiography is a major technology to assist doctors during cardiac interventional surgeries. In Coronary angiography, a radiopaque contrast agent is injected into a blood vessel and X-rays are taken to produce detailed images of the blood vessel. Coronary angiography provides information about the coronary arteries, which supply the heart with oxygen-rich blood. During insertion, the doctor uses fluoroscopy (a continuous x-ray procedure) to observe the progress of the catheter as it is threaded into place. Side effects of radiopaque contrast agents include allergic reactions and kidney damage. To get a coronary angiography video with a high frame rate, the doctor needs to increase the exposure frequency and intensity of the X-ray.  This will inevitably increase the X-ray harm to patients and surgeons. The video frame interpolation is a solution that can synthesize coronary angiography video with a high frame rate without increasing the exposure frequency and intensity of the X-ray.

 Video frame interpolation is mainly to synthesize several frames in the middle of two adjacent frames of the original video. Video frame interpolation is a widely used method in the field of video processing, but never used in coronary angiography videos. Video frame interpolation can be applied to  generate slow motion video, increase video frame rate, and frame recovery in video streaming. The traditional video frame interpolation algorithm mainly includes the following steps: bidirectional motion estimation, motion interpolation, occlusion inference, and motion compensated frame interpolation. However, these methods introduce various disturbances, such as blur, ghost etc. In the past few years, deep-learning methods, especially deep convolutional network methods[2], [3], [7], [12]-[14] have been effectively applied and extended in the field of video frame interpolation. These methods use neural networks to extract kernel functions and optical flow features, and Apply adjacent frames to compose the middle frame. Although these methods have been greatly improved, the effect is still not very good when dealing with  large object motion and occlusion.
 
 The DAIN method[2] has good performance in the field of video frame interpolation. In the method ,a variety of neural network structures are used to extract image context features, optical flow features, depth map features[4][5][6], and applies adaptive  convolutional layers[3] to synthesize new video frames. This method can deal with the situation of adapting to strong motion and large object occlusion. 
 %传统的视频插帧算法主要包括以下步骤双向运动估计，运动插值，遮挡推理，运动补偿帧插值。但是这些方法会引入各种干扰，比如模糊，鬼影等。过去几年深度学习方法尤其是深度卷积网络方法在视频插帧领域得到了有效的应用和扩展，例如Sepconv，Supersolmo，Deep-aware等方法，这些方法通过神经网络提取核函数和光流特征，并应用相邻帧合成新的视频帧。虽然这些方法有了很大的改进，但是在处理快速运动和大物体遮挡情况时，效果依然不是很好。
 
 In this work, we apply video frame interpolation to coronary angiography video  interpolation to increase the frame rate and generate slow motion videos. The video frame interpolation algorithm we choose is the DAIN method. Moreover, we establish a new dataset containing 95,039 triplets where each triplet contains 3 consecutive coronary angiographic frames  to retrain the network model of DAIN.Then we utilize the trained network model to interpolate the coronary angiography videos and analyze the results of interpolation frames. Finally, we compare the performance of different frame interpolation algorithms in different periods of the cardiac cycle.
 
 We make the following contributions in this work:
 
 1. We innovatively apply deep learning-based video frame interpolation algorithms to coronary angiography videos.
 
 2. We make a new coronary angiography dataset for video interpolation algorithm.
 
 %3. We analyze the performance difference of different video frame interpolation algorithms in coronary angiography video, and analyze the performance difference of video frame interpolation algorithms in different periods of heartbeat.
%在心脏介入手术过程中，心脏冠脉造影技术是辅助医生进行手术的主要技术手段.在x光的连续照射下，医生通过导管注入造影剂以实时判断冠脉血管的位置和状态。医生为了准确的判断导管和冠脉血管的状态，有时需要提高造影剂的浓度或者重复注入造影剂。造影剂的使用对患者的肾功能是很大的负担，为了避免过量的造影剂的使用，有时需要把心脏介入手术分阶段进行完成。
%视频插帧主要是通过原始视频的相邻两帧画面合成中间的若干帧画面，这是在视频处理领域被广泛应用的方法。视频插帧可以应用于显示过程的运动模糊和抖动，慢动作合成，以及视频编码和传输后的画面恢复等方向，本文中我们将视频插帧应用于冠脉造影视频的插帧，形成慢动作效果的视频，使医生在手术过程中能够清楚的看到冠脉狭窄和堵塞的状况，或者更清楚的判断支架放入冠脉以后，血液流动是否正常，减少手术中造影剂的使用量。

\section{\label{sec:level1}Related Work}
 Recently deep learning and specifically convolutional neural networks(CNNs) have been successfully applied in  computer vision areas, which inspired various deep learning based frame interpolation methods. The early works on CNN-based video frame interpolation[1] proposed an CNN architecture that takes two input frames and directly estimate the intermediate frame. However, these  typical approaches often lead to blurred outputs.
 
 The later methods, instead of directly computing the output pixels, mainly focused on where to find the pixel from input frames and estimated spatially-adaptive interpolation kernels to synthesize pixels from a large neighborhood. In the AdaConv method[13], the adaptive filter is convolved with adjacent input frame images to synthesize the data of the middle frame. But this method can only deal with moving objects within a size of 41x41 pixels at most which can not meet the situation of strenuous exercise and occlusion. A separable neural network calculation method is applied in the SepConv[12] method, which can greatly save the use of physical memory during the frame insertion process and shorten the time used to train the network.

%在 SepConv方法中没有直接生成像素值，而是通过cnn网络生成了自适应的卷积滤波器，自适应滤波器分别和相邻输入帧图像进行卷积处理，以合成中间一帧的数据。但是这种方法最多只能处理51x51像素大小的运动物体。不能满足剧烈运动和和遮挡的情况。在AdaConv方法中应用了一种可分离神经网络的计算方法，能够大量的节省插帧过程中物理内存的使用，缩短了训练网络所用的时间。

In the SuperSlomo method[7], a CNN is used for bidirectional motion estimate, then a simplified motion interpolation method is applied, and at the end a second CNN performs motion estimate refinement and occlusion reasoning. This work achieved overwhelming quality when applied on videos taken at high frame-rates. However, it seems that they do not aim at covering a wide range of motions.

The adaptive deformable neural network[19] is defined in creative deformable convolution. This method trains a weight kernel at each pixel value position, and then each kernel weight corresponds to a pixel value with an offset. Any positive number, so the bi-linear interpolation method is used to synthesize the pixel value at any position. It performs better under intense movement and occlusion

In the DAIN method[2], a pretrained model is used to extract the context feature map[14], depth features map[8], [9], frame interpolation kernel, and optical flow feature map[11] of the two adjacent frames, and the adaptive convolution layer is used to synthesize any frame data between two adjacent frames. The DAIN method can synthesize video frames at arbitrary positions and has a good performance in dealing with the situation of strong motion and occlusion of objects.

\section{\label{sec:level1}Method}
In this section, we rely on the DAIN method proposed by Bao.w et al.[2]. We first  introduce the DAIN method and describe the main architecture of the network.  And then we redefine the loss function and introduce the implementation details.

\begin{figure*}[h]
	\centering
	\includegraphics[width=16cm,height=7cm]{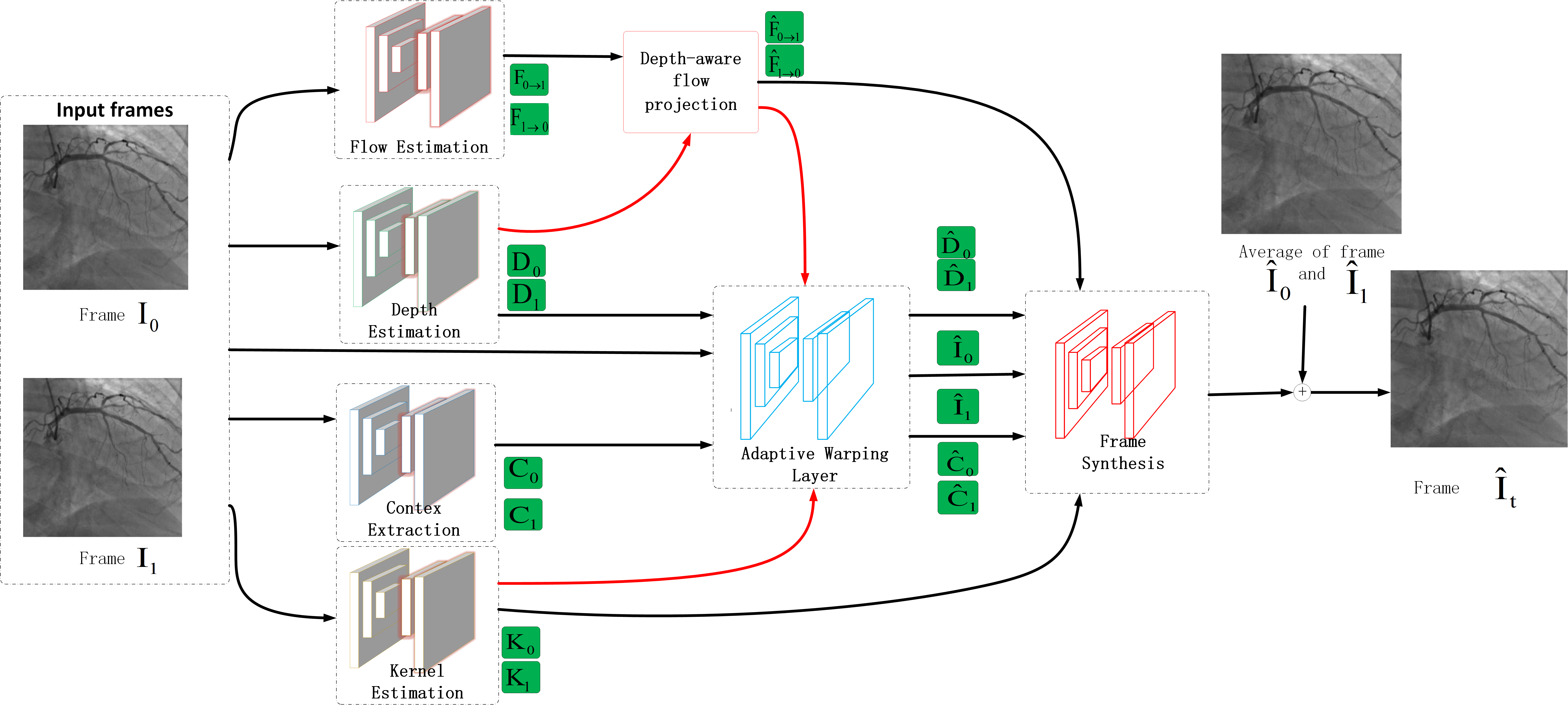}
	
	\caption{ Architecture of DAIN model[2]. The model consists of the following submodules: the flow estimation, depth estimation, context extraction, kernel estimation, Depth-ware flow projection layer, adaptive warping layer, and frame synthesis networks. }
\end{figure*}

\subsection{\label{sec:level2}Depth-Aware Video Interpolation }
The main architecture of the DAIN model is shown in Fig.1. Given two input images $ I_0$ 
and $I_1$ and a time $ t \in (0,1)$,  the goal is to predict the interpolate image $\hat{I}_t$ at 
time $T=t$. The DAIN model consists of the following submodules: the flow estimation, depth estimation,
 context extraction, kernel estimation, and frame synthesis networks. The DAIN algorithm  uses the
  proposed depth-aware flow projection layer to obtain intermediate flows and then warps the input
   frames,depth maps, and contextual features within the adaptive warping layer. Finally,
    the frame synthesis network generates the output frame with residual learning.

The DAIN model utilizes the pretrained PWC-Net[11]
 as the optical flow estimation network. $F_{0 \to 1} $ 
 and   $F_{1 \to 0}$  are the estimated optical flows 
 by the flow estimation network . $F_{0 \to 1} $ and
  $F_{1 \to 0}$  denote the optical flow from $I_0$ to
   $I_1$ and$I_1$ to $I_0$ ,respectively. The depth 
   estimation network use the hourglass architecture[9],
    [16]. $D_0$ and $D_1$ are the estimated depth map of
     $I_0$ and $I_1$. The DAIN method contains a 
     depth-aware flow projection[2] which combine 
     the optical flow and the depth map to get the
      projected flow at time $T=t$ . The projected 
      flow $F_{t \to 0}$ can be obtained from the 
      flow $F_{0 \to 1}$ and the depth map $D_0$. 
      The projected flow $F_{t \to 1}$ can 
      be obtained from the flow $F_{1 \to 0}$ and the depth map $D_1$.

The projected flow $F_{t \to 1} $ is defined by:
\begin{equation}\label{key}
F_{t \to 1}(x)=-(1-t)\cdot{\frac{\sum\limits_{y\in\mathcal{S}(x) } \frac{1}{{D}_{1}(y)}\cdot F_{1 \to 0}(y)}{\sum\limits_{y\in\mathcal{S}(x)}\frac{1}{{D}_{1}(y)}}}
\end{equation}
\begin{equation}\label{key}
\mathcal{S}(x) = \left\{y:round(y+(1-t)F_{1\to0}(y)) =x, \forall y \in [H,W] \right\}
\end{equation}
Where $\mathcal{S}(x)$ indicates the set of pixels that pass through the position  $x$ at time $t$. 
The projected flow $F_{t \to 0} $ is defined by:
\begin{equation}\label{key}
F_{t \to 0}(x)=-t\cdot{\frac{\sum\limits_{y\in\mathcal{S}(x) } \frac{1}{{D}_{0}(y)}\cdot F_{0 \to 1}(y)}{\sum\limits_{y\in\mathcal{S}(x)}\frac{1}{{D}_{0}(y)}}}
\end{equation}

\begin{equation}\label{key}
\mathcal{S}(x)=\left\{y:round(y+tF_{0\to1}(y)) = x, \forall y \in  [H,W] \right\}
\end{equation}
The context extraction model consists a pretrained ResNet[17]. The contextual features $C_0$ and $C_1$ are extracted from the input frames $I_0$ and $I_1$.The interpolation kernels $K_0$ and $K_1$ are estimated by a U-Net architecture[15] network. With the interpolation kernels ($K_0$ and $K_1$) and interpolated flows($F_{t\to0}$ and $F_{t\to1}$) generated from the depth-aware flow projection layer, using the adaptive warping layer[3] to warp the input frames( $I_0$ and $I_1$), depth maps($D_0$ and $D_1$),contextual features($C_0$ and $C_1$). To generate the output frame$\widehat{I}_t $, the algorithm utilizes a frame synthesis network, which consists of 3 residual blocks. The algorithm concatenates  projected flows( $\widehat{F}_{t \to 0}$ and $\widehat{F}_{t \to 1}$ ), and interpolation kernels($K_0$ and $K_1$),the warped input frames($\widehat{I}_0$ and $\widehat{I}_1$), warped depth maps($\widehat{D}_0$ and $\widehat{D}_1$), warped contextual features($\widehat{C}_0$ and $\widehat{C}_1$) as the input to the frame synthesis network. In addition, the algorithm linearly blend the two warped frames($\widehat{I}_0$ and $\widehat{I}_1$) and enforce the network to predict the residuals between the ground-truth frame and the blended frame.

%The projected flow ${F}_{t \to 1}$ is defined by:

%where the weight $ {w}_{1}$ is the reciprocal of depth:
%\begin{equation}\label{key}
%{w}_{1}(y)=\frac{1}{{D}_{1}(y)}
%\end{equation}
%As shown in Fig.1 ,the depth-aware flow projection combine depth values and generate the flow vector pointing to the closer pixel, instead of obtaining an average flow vector.

%On the other hand, there may exist positions where none of the flow vectors pass through, leading to holes in the intermediate flow. To fill in the holes, the algorithm use the outside-in strategy : the flow in the hole position is computed by averaging the available flows from its neighbors:
%\begin{equation}\label{key}
%F_{t \to 0}(x)=\frac{1}{|\mathcal{N}(x)|}\sum\limits_{x^{'}\in\mathcal{N}(x)}F_{t \to 0}(x^{'})
%\end{equation}

%\subsection{\label{sec:level2}Video Frame Interpolation}
\subsection{\label{sec:level3}Implementation Details}
Loss Function. We denote the synthesized frame by$ \hat{I}_{t}$
and the ground-truth frame by $ I_{t}$. We retrain the DIAN 
model by optimizing the following loss function:

\begin{equation}\label{key}
       L={   \lambda_1l_1+\lambda_2l_2}
\end{equation}

where $\lambda_1$ = 0.95 and $\lambda_2$ = 0.05.

The loss $l_1$ models how good the reconstruction fo the intermediate frame is:

\begin{equation}\label{key}
l_1=\sum\limits_x\rho||\hat{I}_{t}(x)-I_{t}(x)||_1
\end{equation}

where $\rho(x)=\sqrt{x^{2}+\epsilon^{2}}$is the Charbonnier Function. We set the constant $\epsilon$ to $1e-4$.

The purpose of coronary angiography video frame interpolation is to obtain a clearer coronary artery structure. So we  define the structure loss $l_2$ as 

\begin{equation}\
l_2=\sum\limits_x\rho||\phi{\hat{I}}_{t}(x)-\phi I_{t}(x)||_1
\end{equation}

where $\phi$ denote the conv4\_x features of a pretrained ResNet[17].

Training Dataset. In  order to  train a better network model conforming to coronary angiography scene we create a new training dataset. The dataset contains 950,39 triplets of consecutive frames with a resolution of  480x360. The triplets are extracted from 31 coronary angiography videos ，the total duration of which is about 25 hours. The videos are produced by recording screen images during cardiac intervention. We processed these videos and selected suitable coronary angiography fragments as the data set.  The image frame rate of coronary angiography is 7-15fps, but the frame rate used for recording screen images is 25fps, so we  preprocess the recorded videos to remove duplicate video frames. Coronary angiography images contain some patient's information, so we  occlude and remove these information to ensure that the patient's privacy is not leaked. Also we augument the training data by horizontal and vertical flipping.

%训练数据集。 为了训练符合冠状动脉造影场景的更好的网络模型，我们创建了一个大型训练数据集。 数据集包含95039张三连串的连续帧，分辨率为480x360。 从30个冠状动脉造影视频中提取三联体，其总持续时间约为25小时。 这些视频是通过在冠状动脉造影过程中记录屏幕图像来制作的，因此这些视频记录了手术期间的所有器械图像，包括冠状动脉造影术的视频。 我们处理了这些视频，并选择了合适的冠状动脉造影片段作为数据集。 冠状动脉造影的图像帧率为7-15fps，但用于记录屏幕图像的帧率为25fps，因此我们必须对记录的视频进行预处理以删除重复的视频帧。 冠状动脉造影图像包含一些患者信息，因此我们已遮挡并删除了这些信息，以确保不会泄露患者的隐私。
Training strategy. Every triplet in the dataset  contains 3 consecutive video frames, which are the previous frame, the middle frame, and the next frame. we train the network to predict the middle frame that serves as ground truth. We use a pre-trained model as the initial model for training. The learning rate is initially 0.0001 and decays half every 20 epochs. The batch size is 4 and the the network is trained for 100 epochs. We train the network on an NVIDIA 2080Ti GPU card,which takes about 30 hours.

\section{\label{sec:level1}Experiments and Results}
In this section, we first introduce the testing datasets. Second, we compare the output of the pretrained model with the retrained DAIN model. Then we compare our results with stat-of-the-art frame interpolation approaches and analyze the performance of different frame interpolation methods in different cardiac cycles. Finally we discuss how to further improve the current results.
\subsection{\label{sec:level2}Testing  Dataset }
 We randomly selected 1,000 triplets  from the coronary angiography dataset as one test dataset which is represented by D1000. At the same time, we extracted four coronary angiography video clips, each of which contained  30-45 continuously frames, and each video clip includes the process of injection, diffusion, and disappearance of the contrast agent. We use VC1, VC2, VC3, VC4 to represent these four  video clips. For these four clips, we use odd frames to predict even frame, and then use even frames to predict odd frame, so we can compare all the  predicted frames  with the original frames. We utilize the PSNR and SSIM values as the evaluation criteria for the two-frame image error.   
%实验数据及评估标准：我们随机的从冠脉造影数据集抽取1000组。每个图像的的前一帧图像和后一帧图像作为输入，预测中间帧。预测帧与真实的中间帧进行比较，计算两张图像之间的误差。同时我们抽取四个冠脉造影片段，每个片段包含30-45帧图像，每一个片段内容包括注入造影剂，造影剂扩散，造影剂消失的过程。对于这四个片段，我们用奇数帧预测偶数帧数据，然后再用偶数帧预测奇数帧的数据，这样我们就可以用所有的预测帧与原始帧进行质量对比和评价。我们采用PSNR值和SSIM值作为两帧图像误差的评估标准。
\begin{table}[tbp]      
	%h：hear，t：top，b：bottom，p：page，下一页。
	\centering
	\caption{ The  experiments results   of different datasets.}
	\label{tab1}
	\begin{tabular}{c c c c cccc}
		\hline
		Test data          & D1000        & VC1      & VC2                               \\ 
		\hline  
		& SSIM,PSNR     & SSIM,PSNR     & SSIM,PSNR       \\
		\hline
		DAIN-Pre      &0.961 40.142 & 0.902 33.839         &  0.900 34.368          \\
		DAIN-Re            &0.962 40.523 & 0.904 34.068         &  0.902 34.754         \\
		\hline
		
	\end{tabular}
	%\caption{The  experiments results   of different datasets }
\end{table}

\begin{figure}[]
	\begin{minipage}[t]{0.47\linewidth}%并排放两张图片，每张占行的0.4，下同 
		\centering     %插入的图片居中表示
		\includegraphics[width=1.00\textwidth]{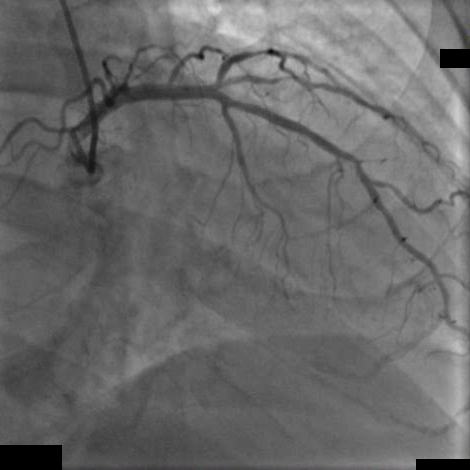}
		%\caption{Original frame $I_0$ }%图片的名称
		Original frame $I_0$ 
		\label{}%标签，用作
	\end{minipage} 
	\hfill
	\begin{minipage}[t]{0.47\linewidth}
		\centering
		\includegraphics[width=1.00\textwidth]{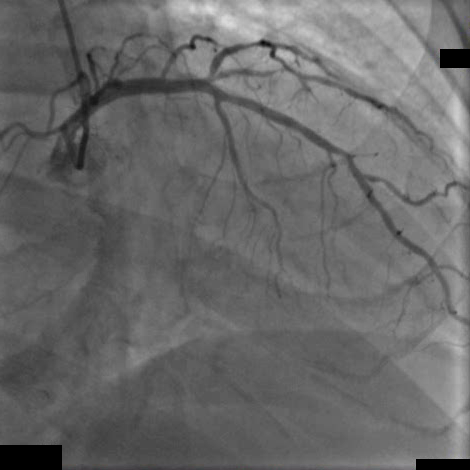}
		%\caption{Original frame $I_1 $ }%图片的名称
		Original frame $I_1 $
		\label{}
	\end{minipage}
	\hfill
	
	\begin{minipage}[t]{0.32\linewidth}%并排放两张图片，每张占行的0.4，下同 
		\centering     %插入的图片居中表示
		\includegraphics[width=1.0\textwidth]{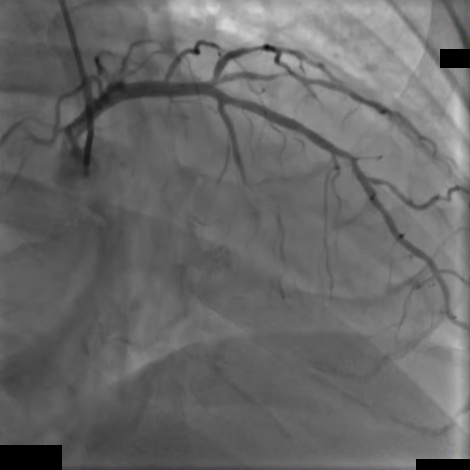}
		%\caption{ $I^{Pre}_{t=0.5}$,PSNR=34.911}%图片的名称
		$I^{Pre}_{t=0.5}$,PSNR=34.91
		\label{$I^{Pre}_{t=0.5}$,PSNR=34.91}%标签，用作
	\end{minipage} 
	\hfill
	\begin{minipage}[t]{0.32\linewidth}%并排放两张图片，每张占行的0.4，下同 
		\centering     %插入的图片居中表示
		\includegraphics[width=1.0\textwidth]{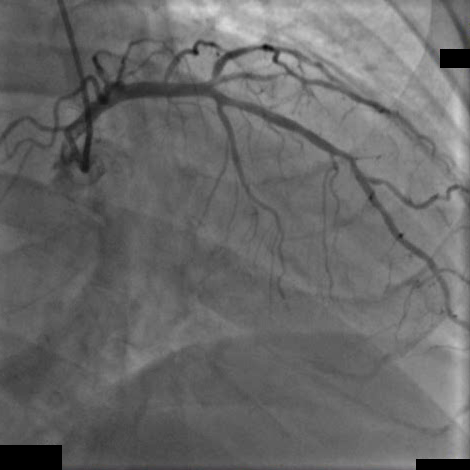}
		%\caption{Ground truth }%图片的名称
		Ground truth
		\label{}%标签，用作
	\end{minipage} 
	\hfill
	\begin{minipage}[t]{0.32\linewidth}%并排放两张图片，每张占行的0.4，下同 
		\centering     %插入的图片居中表示
		\includegraphics[width=1.0\textwidth]{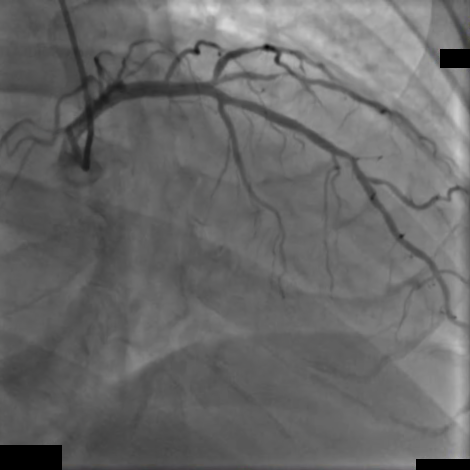}
		%\caption{$I^{Re}_{t=0.5}$,PSNR=35.280  }%图片的名称
		$I^{Re}_{t=0.5}$,PSNR=35.28
		\label{}%标签，用作
	\end{minipage} 
	\hfill
	%\caption[Fig.1]{}

	\caption{  Comparison of the output of the retrained model and the pretrained model. The  $I^{Pre}_{t=0.5}$ is the predicted video frame of pretrained model and the $I^{Re}_{t=0.5}$ is the predicted video frame of retrained model  }
\end{figure}

\begin{figure}[]
	
	\begin{tikzpicture}
	\begin{axis}[
	xlabel={Index of frames},
	ylabel={PSNR【[dB]},
	xmin=0, xmax=30,
	ymin=30, ymax=42,
	xtick={0,5,10,15,20,25,30},
	ytick={30,32,34,36,38,40,42},
	legend pos=north east,
	ymajorgrids=true,
	grid style=dashed,
	]
	
	\addplot[
	color=blue,
	mark=square,
	]
	table {./20200107/20200202/DAIN1/1_4.txt};
	%\legend{slomo}
	\addlegendentry{DAIN-Pre}
	
	\addplot[
	color=red,
	mark=square,
	]
	table {./20200107/20200202/DAIN1/2_4.txt};
	%\legend{Dain}
	\addlegendentry{DAIN-Re}

	\end{axis}
	
	\end{tikzpicture}
	\caption{  PSNR of interpolation frame  changes with frame index, the test data is VC4}
\end{figure}
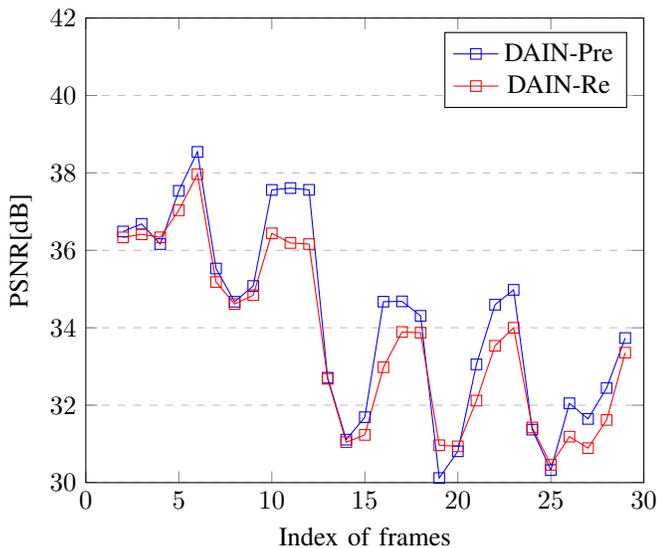
\subsection{\label{sec:level2}Results  }
We use the testing dataset  to compare the results of the pretrained DAIN model with the retrained model. Fig.2 shows the inputs of two original frames, the original middle frame as ground truth, the interpolation midldle frame of the pretrained model, and interpolation  midldle frame of the retrained model. The PSNR of  $I^{Re}_{t=0.5}$ is 0.3dB higher than  $I^{Pre}_{t=0.5}$. The comparison of the average values of PSNR and SSIM for D1000  and two video clips is recorded in Table\uppercase\expandafter{\romannumeral1}. It shows the retrained model  has higher PSNR and SSIM than the pretrained model. Fig.3 shows the PSNR changing with frame index of one coronary angiography video clip. The retrained model has  higher PSNR on every frame. From Fig.3 we can see that the PSNR value has a similar periodic change, which is related to the cardiac cycle. We will analyze the reasons for this phenomenon in the subsequent results. 

\begin{figure*}[]
	\begin{minipage}[t]{0.195\linewidth}%并排放两张图片，每张占行的0.4，下同 
		\centering     %插入的图片居中表示
		%\title{DIAN}\
		%DAIN
		\includegraphics[width=1.0\textwidth]{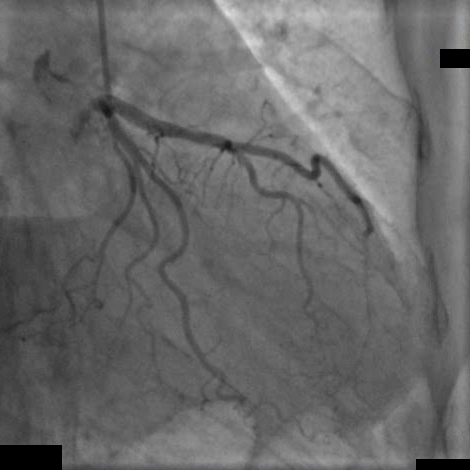}
		
		%	\hfill
		%\caption{Frame $I_0$ }%图片的名称
		%\label{DAIN}%标签，用作
	\end{minipage} 
	\hfill	
	\begin{minipage}[t]{0.195\linewidth}%并排放两张图片，每张占行的0.4，下同 
		\centering     %插入的图片居中表示
		%\title{DIAN}\
		%DAIN
		\includegraphics[width=1.00\textwidth]{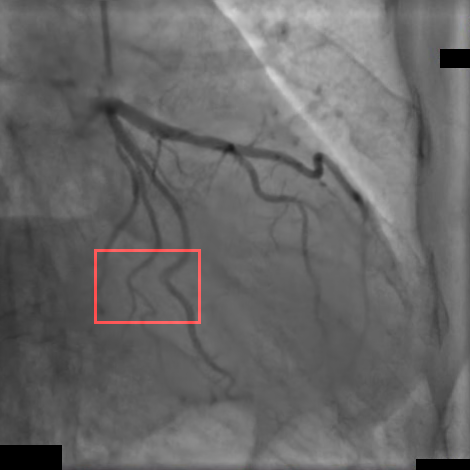}
		
		%	\hfill
		%\caption{Frame $I_0$ }%图片的名称
		%\label{DAIN}%标签，用作
	\end{minipage} 
	\hfill
	\begin{minipage}[t]{0.195\linewidth}%并排放两张图片，每张占行的0.4，下同 
		\centering     %插入的图片居中表示
		\includegraphics[width=1.0\textwidth]{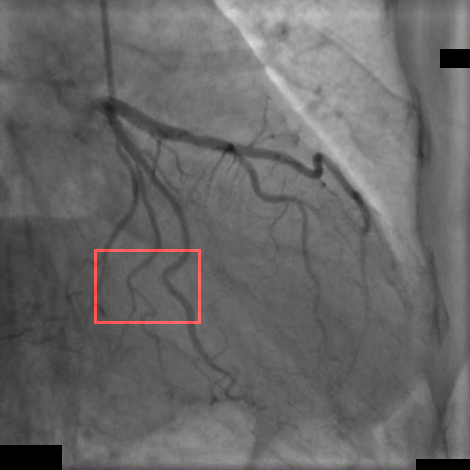}
		%	\caption{Frame $I_0$ }%图片的名称
		%\label{}%标签，用作
	\end{minipage}
	\hfill
	\begin{minipage}[t]{0.195\linewidth}%并排放两张图片，每张占行的0.4，下同 
		\centering     %插入的图片居中表示
		\includegraphics[width=1.0\textwidth]{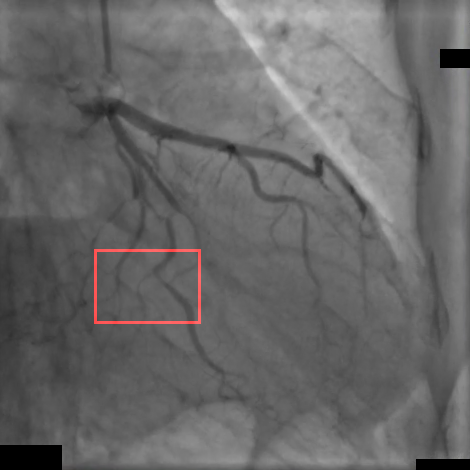}
		%	\caption{Frame $I_0$ }%图片的名称
		%\label{}%标签，用作
	\end{minipage} 
	\hfill
	\begin{minipage}[t]{0.195\linewidth}%并排放两张图片，每张占行的0.4，下同 
		\centering     %插入的图片居中表示
		\includegraphics[width=1.0\textwidth]{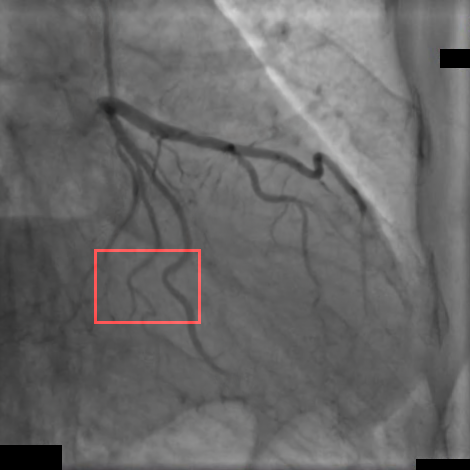}
		%\title{original frame }
		%	\caption{Frame $I_0$ }%图片的名称frame
		%\label{}%标签，用作
	\end{minipage} 
	\hfill
		\begin{minipage}[t]{0.195\linewidth}%并排放两张图片，每张占行的0.4，下同 
		\centering     %插入的图片居中表示
		%\title{DIAN}\
		%DAIN
		\includegraphics[width=1.0\textwidth]{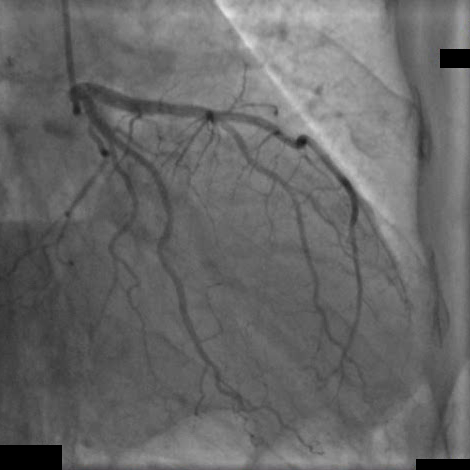}
		
		%	\hfill
		%\caption{Frame $I_0$ }%图片的名称
		%\label{DAIN}%标签，用作
	\end{minipage} 
	\hfill
	\begin{minipage}[t]{0.195\linewidth}%并排放两张图片，每张占行的0.4，下同 
		\centering     %插入的图片居中表示
		\includegraphics[width=1.0\textwidth]{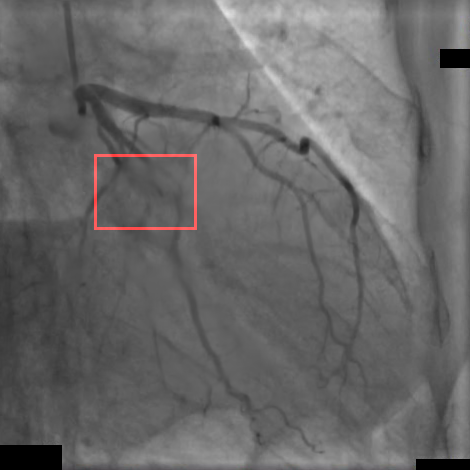}
		%	\caption{Frame $I_0$ }%图片的名称
		%\label{}%标签，用作
	\end{minipage} 
	\hfill
	\begin{minipage}[t]{0.195\linewidth}%并排放两张图片，每张占行的0.4，下同 
		\centering     %插入的图片居中表示
		\includegraphics[width=1.0\textwidth]{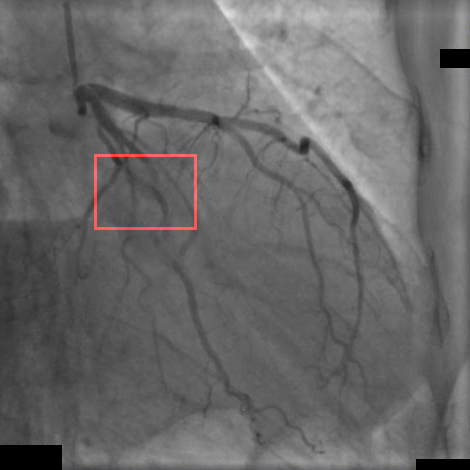}
		%	\caption{Frame $I_0$ }%图片的名称
		%\label{}%标签，用作
	\end{minipage} 
	\hfill 
	\begin{minipage}[t]{0.195\linewidth}%并排放两张图片，每张占行的0.4，下同 
		\centering     %插入的图片居中表示
		\includegraphics[width=1.0\textwidth]{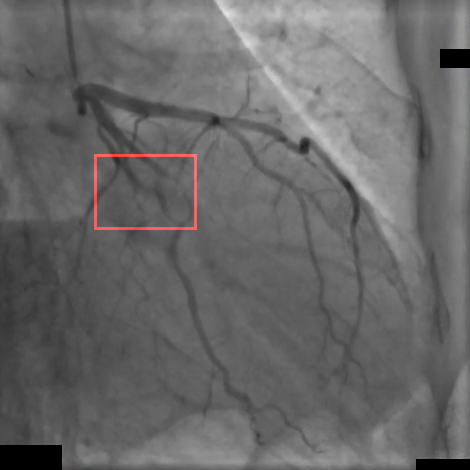}
		%	\caption{Frame $I_0$ }%图片的名称
		%\label{}%标签，用作
	\end{minipage}
	\hfill 
	\begin{minipage}[t]{0.195\linewidth}%并排放两张图片，每张占行的0.4，下同 
		\centering     %插入的图片居中表示
		\includegraphics[width=1.0\textwidth]{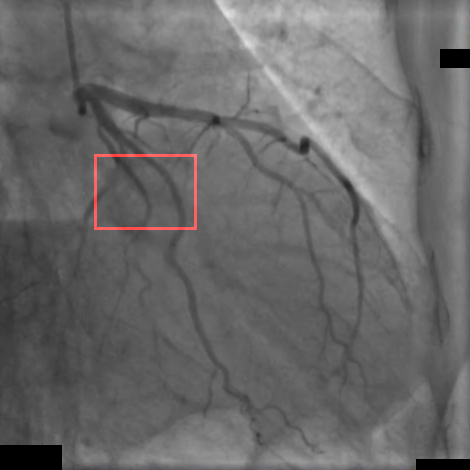}
		%	\caption{Frame $I_0$ }%图片的名称
		%\label{}%标签，用作
	\end{minipage} 
	\hfill
		\begin{minipage}[t]{0.195\linewidth}%并排放两张图片，每张占行的0.4，下同 
		\centering     %插入的图片居中表示
		%\title{DIAN}\
		%DAIN
		\includegraphics[width=1.0\textwidth]{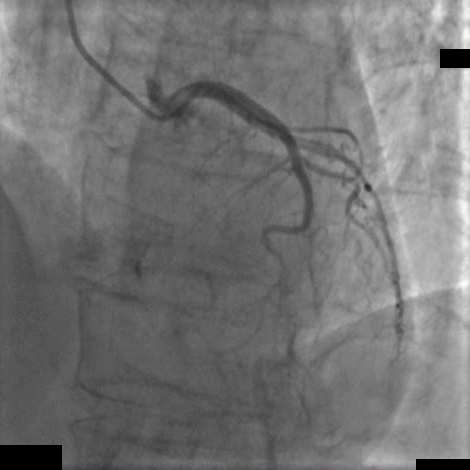}
		
		%	\hfill
		%\caption{Frame $I_0$ }%图片的名称
		%\label{DAIN}%标签，用作
	\end{minipage} 
	\hfill
	\begin{minipage}[t]{0.195\linewidth}%并排放两张图片，每张占行的0.4，下同 
		\centering     %插入的图片居中表示
		\includegraphics[width=1.0\textwidth]{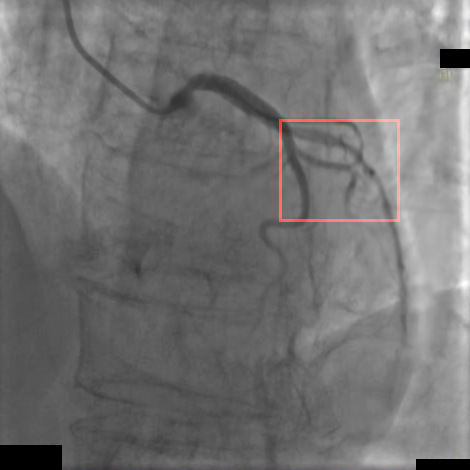}
		%	\caption{Frame $I_0$ }%图片的名称
		%\label{}%标签，用作
	\end{minipage} 
	\hfill
	\begin{minipage}[t]{0.195\linewidth}%并排放两张图片，每张占行的0.4，下同 
		\centering     %插入的图片居中表示
		\includegraphics[width=1.0\textwidth]{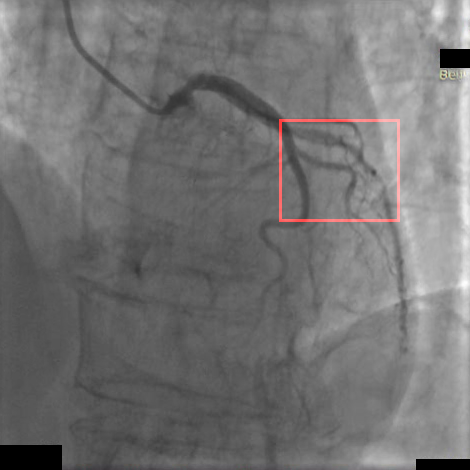}
		%	\caption{Frame $I_0$ }%图片的名称
		%\label{}%标签，用作
	\end{minipage} 
	\hfill
	\begin{minipage}[t]{0.195\linewidth}%并排放两张图片，每张占行的0.4，下同 
		\centering     %插入的图片居中表示
		\includegraphics[width=1.0\textwidth]{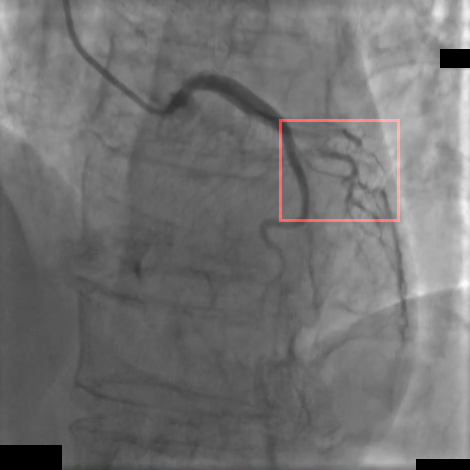}
		%	\caption{Frame $I_0$ }%图片的名称
		%\label{}%标签，用作
	\end{minipage} 
	\hfill
	\begin{minipage}[t]{0.195\linewidth}%并排放两张图片，每张占行的0.4，下同 
		\centering     %插入的图片居中表示
		\includegraphics[width=1.0\textwidth]{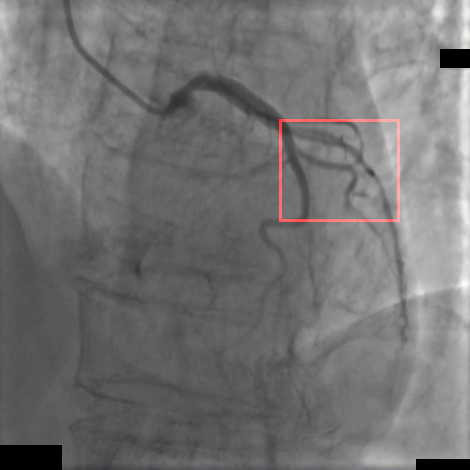}
		%	\caption{Frame $I_0$ }%图片的名称
		%\label{}%标签，用作
	\end{minipage} 
	\hfill
		\begin{minipage}[t]{0.195\linewidth}%并排放两张图片，每张占行的0.4，下同 
		\centering     %插入的图片居中表示
		%\title{DIAN}\
		%DAIN
		\includegraphics[width=1.0\textwidth]{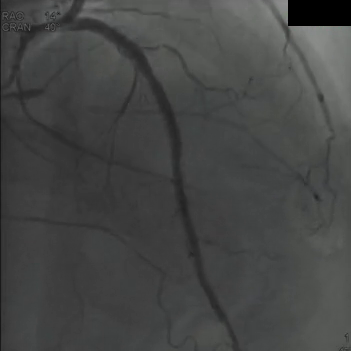}
		
		%	\hfill
		%\caption{Frame $I_0$ }%图片的名称
		%\label{DAIN}%标签，用作
	\end{minipage} 
	\hfill
	\begin{minipage}[t]{0.195\linewidth}
		\centering
		\includegraphics[width=1.0\textwidth]{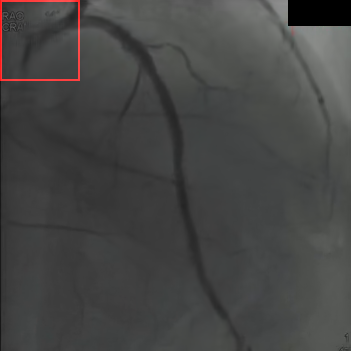}
		%\caption{SepConv-l1}%图片的名称
		%	\label{}
	\end{minipage}
	\hfill		
	\begin{minipage}[t]{0.195\linewidth}
		\centering
		\includegraphics[width=1.0\textwidth]{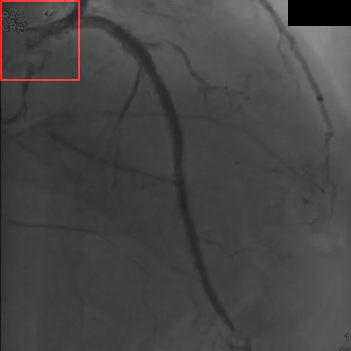}
		%\caption{SepConv-lf }%图片的名称
		%	\label{}
	\end{minipage}
	\hfill			
	\begin{minipage}[t]{0.195\linewidth}
		\centering
		\includegraphics[width=1.0\textwidth]{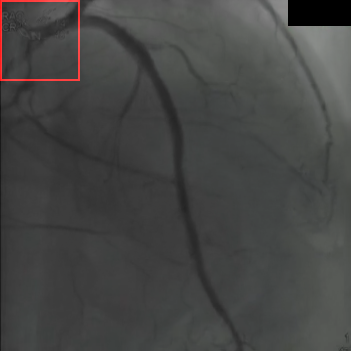}
		%\caption{SuperSlomo }%图片的名称
		%	\label{}
	\end{minipage}
	\hfill	
	\begin{minipage}[t]{0.195\linewidth}
		\centering
		\includegraphics[width=1.0\textwidth]{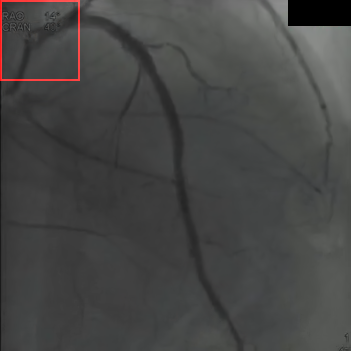}
		%\caption{DAIN }%图片的名称
		%	\label{}
	\end{minipage}
	\hfill
\begin{minipage}[t]{0.19\linewidth}
	\centering
	%\includegraphics[width=1.11\textwidth]{./20200107/20200202/sepconv/l11/52/029-1.png}
	%\caption{SepConv-l1}%图片的名称
	%	\label{}
	Ground truth
\end{minipage}
\hfill
	\begin{minipage}[t]{0.19\linewidth}
	\centering
	%\includegraphics[width=1.11\textwidth]{./20200107/20200202/sepconv/l11/52/029-1.png}
	%\caption{SepConv-l1}%图片的名称
	%	\label{}
	 SepConv-l1
\end{minipage}
\hfill		
\begin{minipage}[t]{0.19\linewidth}
	\centering
	%\includegraphics[width=1.11\textwidth]{./20200107/20200202/sepconv/lf1/52/029-1.png}
	%\caption{SepConv-lf }%图片的名称
	%	\label{}
	 SepConv-lf
\end{minipage}
\hfill			
\begin{minipage}[t]{0.19\linewidth}
	\centering
	%\includegraphics[width=1.11\textwidth]{./20200107/20200202/superslomo/52/029-1.png}
	%\caption{SuperSlomo }%图片的名称
	%	\label{}
	 SuperSlomo
\end{minipage}
\hfill	
\begin{minipage}[t]{0.19\linewidth}
	\centering
	%\includegraphics[width=1.11\textwidth]{./20200107/20200202/DAIN1/52/029-1.png}
	%\caption{DAIN }%图片的名称
	%	\label{}
	DAIN
\end{minipage}
\hfill
	\hfill	
	\caption{The interpolation Frame of  DAIN , SepConv-l1 ,SepConv-lf ,SuperSlomo. It mainly shows the performance of several methods in the case of occlusion and large object movement. As can be seen from the figure, the DAIN method performs better in both aspects. }	
	
\end{figure*}

 \begin{table*}[]      
	%h：hear，t：top，b：bottom，p：page，下一页。
	\centering
	\caption{ The average PSNR value and SSIM value of several frame interpolation algorithms.}
	\label{tab2}
	\begin{tabular}{cccccc}
		\hline
		Test data          & VC1        & VC2      &VC3   &VC4        & D1000                              \\ 
		\hline  
		&SSIM , PSNR    &SSIM , PSNR     &SSIM , PSNR    &SSIM , PSNR    &SSIM , PSNR \\
		\hline
		SepConv-l1      &0.901 33.657       &0.901 34.672 &   0.904 33.649      &0.902 33.745     & 0.962 39.823 \\
		SepConv-lf      &0.878 32.768        &0.883 34.039 &0.883 33.003      & 0.883 32.872   &0.956 39.741\\
		SuperSlomo     &0.901 33.838      &0.898 34.468 &     0.902 33.768    &0.901 33.704   &0.958 38.868 \\
		DAIN            & 0.904 34.068         &  0.902 34.754      &0.908 33.977  &0.908 34.205 &0.962 40.523   \\
		\hline
	\end{tabular}
\end{table*}

\begin{figure*}[]
	\begin{minipage}[t]{0.49\linewidth}%并排放两张图片，每张占行的0.4，下同 
		\centering     %插入的图片居中表示
		\begin{tikzpicture}
		\begin{axis}[
		xlabel={Index of frames},
		ylabel={PSNR【[dB]},
		xmin=0, xmax=35,
		ymin=28, ymax=48,
		xtick={0,5,10,15,20,25,30,35,40,45},
		ytick={28,30,32,34,36,38,40,42,44,46,48},
		legend pos=north east,
		ymajorgrids=true,
		grid style=dashed,
		]
		\centering
				\addplot[
		color=red,
		mark=square,
		]
		table {./20200107/20200202/DAIN1/2.txt};
		%\legend{Dain}
		\addlegendentry{DAIN}
		\addplot[
		color=blue,
		mark=square,
		]
		table {./20200107/20200202/superslomo/2.txt};
		%\legend{slomo}
		\addlegendentry{SuperSlomo}
		\addplot[
		color=green,
		mark=square,
		]
		table {./20200107/20200202/sepconv/l11/2.txt};
		%#\legend{Da}
		\addlegendentry{SepConv-l1}
		\addplot[
		color=black,
		mark=square,
		]
		table {./20200107/20200202/sepconv/lf1/2.txt};
		
		\addlegendentry{SepConv-lf}
		\addplot[color=black,mark=ball,]
		coordinates {(6.8,0)(6.8,40)};
		
		\addplot[color=red,mark=ball,]
		coordinates {(11.8,0)(11.8,40)};
		
		\addplot[color=black,mark=ball,]
		coordinates {(18.8,0)(18.8,40)};
		\addplot[color=red,mark=ball,]
		coordinates {(23.8,0)(23.8,40)};
		
		\addplot[color=black,mark=ball,]
		coordinates {(29.8,0)(29.8,40)};
		
		\end{axis}
		\end{tikzpicture}
		%	\caption{Frame $I_0$ }%图片的名称
		%\label{}%标签，用作
	\end{minipage} 
\hfill
	\begin{minipage}[t]{0.49\linewidth}%并排放两张图片，每张占行的0.4，下同 
		\centering     %插入的图片居中表示
		\begin{tikzpicture}
		\begin{axis}[
		xlabel={Index of frames},
		ylabel={PSNR [dB]},xmin=0, xmax=45,ymin=28, ymax=48,
		xtick={0,5,10,15,20,25,30,35,40,45},
		ytick={28,30,32,34,36,38,40,42,44,46,48},
		legend pos=north east,
		ymajorgrids=true,
		grid style=dashed,
		]
		\addplot[
		color=red,
		mark=square,
		]
		table {./20200107/20200202/DAIN1/3.txt};
		%\legend{Dain}
		\addlegendentry{DAIN}
		\addplot[
		color=blue,
		mark=square,
		]
		table {./20200107/20200202/superslomo/3.txt};
		%\legend{slomo}
		\addlegendentry{SuperSlomo}
		\addplot[
		color=green,
		mark=square,
		]
		table {./20200107/20200202/sepconv/l11/3.txt};
		%#\legend{Da}
		\addlegendentry{SepConv-l1}
		
		\addplot[
		color=black,
		mark=square,
		]
		table {./20200107/20200202/sepconv/lf1/3.txt};
		%#\legend{Da}
		\addlegendentry{SepConv-lf}
		
		\addplot[
		color=red,
		mark=ball,
		]
		coordinates {
			(11.5,0)(11.5,40)};
		\addplot[
		color=black,
		mark=ball,
		]
		coordinates {
			(18.5,0)(18.5,40)};
		\addplot[
		color=red,
		mark=ball,
		]
		coordinates {
			(23.5,0)(23.5,40)};
		\addplot[
		color=black,
		mark=ball,
		]
		coordinates {
			(29.5,0)(29.5,40)};
		\addplot[
		color=red,
		mark=ball,
		]
		coordinates {
			(34.5,0)(34.5,40)};
		\addplot[
		color=black,
		mark=ball,
		]
		coordinates {
			(40.5,0)(40.5,40)};
		\end{axis}
		\end{tikzpicture}
		%	\caption{Frame $I_0$ }%图片的名称
		%\label{}%标签，用作
	\end{minipage} 
	\hfill
	\begin{minipage}[t]{0.499\linewidth}
	\centering
	%\includegraphics[width=1.11\textwidth]{./20200107/20200202/superslomo/52/029-1.png}
	%\caption{SuperSlomo }%图片的名称
	%	\label{}
	(a)
\end{minipage}
\hfill	
\begin{minipage}[t]{0.48\linewidth}
	\centering
	%\includegraphics[width=1.11\textwidth]{./20200107/20200202/DAIN1/52/029-1.png}
	%\caption{DAIN }%图片的名称
	%	\label{}
	(b)
\end{minipage}
	\caption{ The PSNR curves of several frame interpolation algorithms on two continuous coronary angiography clips.(a) shows the results of VC2 and (b) shows the results of VC3. The vertical red line indicates the beginning of systole and the vertical black line indicates the beginning of diastole.}	
\end{figure*}
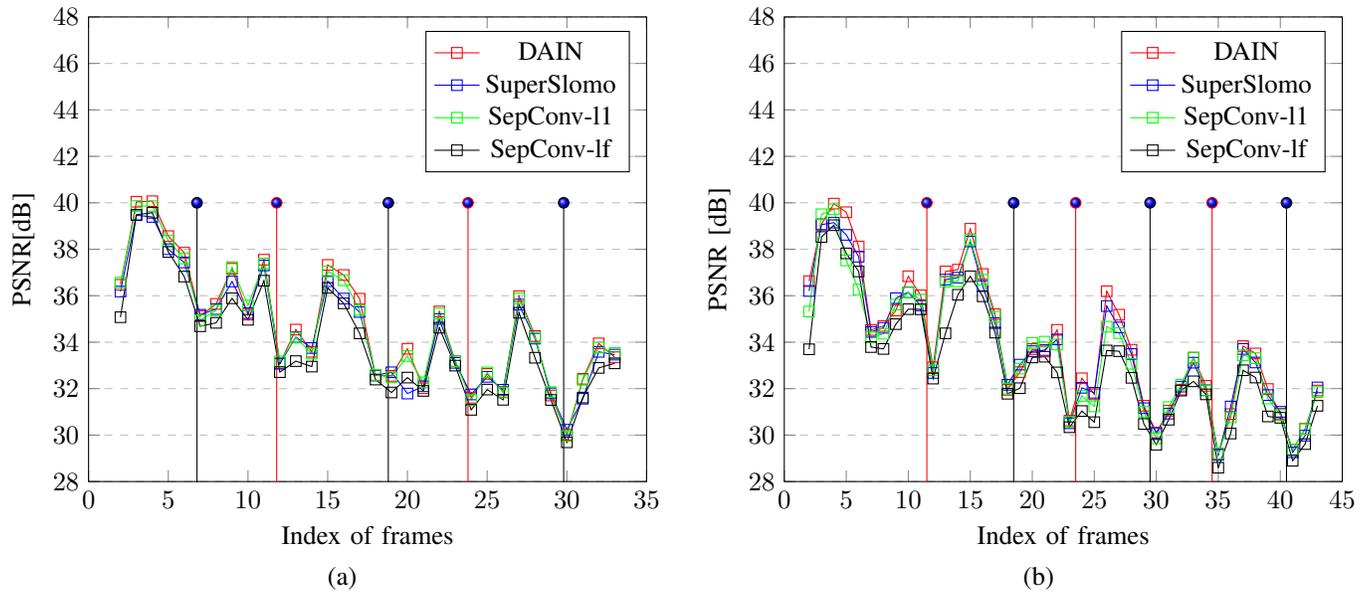

 \subsection{\label{sec:level2} Comparison and Discussion  }
 We compare our results with several stat-of-the-art algorithms including SepConv-l1[12], SepConv-lf[12], SuperSlomo[7]. In order to exclude the influence of the training dataset, we use the coronary angiography dataset to retrain these several network models respectively. Then we use these retrained models to process the testing datasets. 
 
 In Table \uppercase\expandafter{\romannumeral2}, it shows the comparisons of different  methods on the coronary angiography testing datasets. The tables summarizes the average PSNR and SSIM of these experiments. The average PSNR of DAIN method is 0.2dB-1.6dB higher than the other methods. 
 
 In Fig.4, it shows the interpolation frames of several video interpolation methods in the case of occlusion and large object movement. As shown in the figure, the DAIN method performs well in both aspects. The image edges of the interpolated frames are clearer and sharper.
 
 In Fig.5, we plot the PSNR curves of several frame interpolation algorithms on two coronary angiography video clips. First, the overall PSNR curve of DAIN is higher than other methods. Second, all the PSNR  curves has a periodic change. When the systolic and diastolic phases are switched in the cardiac cycle, lower PSNR values appear. We analyze the reason for this result is that when the phase of the cardiac cycle is switched, the optical flow characteristics between two adjacent frames are not obvious enough, and the possible motion vectors of the coronary arteries cannot be accurately predicted. At the same time, we can see that the PSNR value decreases with the index of the frame. The reason for this problem is that the vascular motion area of the coronary angiography image continues to expand with the diffusion of the contrast agent in the coronary arteries. As shown in Figure 5 (b), after the contrast agent has diffused to the end of the coronary artery, the PSNR no longer changes significantly between the 35th and 45th frames. 
 
 In order to further improve the quality of coronary angiography interpolation frame, we can use the trained coronary artery segmentation [18]network model to extract the coronary artery segmentation results from the synthetic frame and the original frame, respectively. We take the error of coronary artery segmentation in two frames as part of the loss function. This method may also improve the problem that the interpolation frames have lower  PSNR values when the cardiac cycle phase switched.
% In Table 3, we calculated the average psnr value and SSIM value of several frame interpolation algorithms at different heartbeat periods, and used them to analyze the differences in performance of these frame interpolation algorithms at different heartbeat stages.

 %\caption{Frame $I_0 , I_{t=0.2} , I_{t=0.4} , I_{t=0.6}, I_{t=0.8}, I_{1}$ }%图片的名称
 %这里我们主要展示了几种方法在有遮挡和大物体移动情况下的表现，从图上可以看出，DAIN方法在这两方面都有很好的表现，合成之后的图像边界更清晰并且拖影现象更少。

\section{Conclusion }
 
In this paper, we innovatively apply the deep learning-based video frame interpolation algorithm to coronary angiography. We establish a new coronary angiography dataset to retrain the network of DAIN method. The retrained model has a better performance in the application scenarios of coronary angiography. Moreover, we  retrain several other deep learning-based algorithms and compare the results of these frame interpolation algorithms. The retained DAIN model has better performance than other methods in the case of occlusion and large object movement. Extensive experiment results demonstrate the feasibility of using the video frame interpolation algorithm to synthesize continuous and clear high frame rate coronary angiography video. With the help of this technology, doctors can significantly reduce exposure frequency and intensity of the X-ray during coronary angiography.
%在这项工作中，我们将基于深度学习的视频帧插值算法应用于冠状动脉造影。我们制作了冠状动脉造影数据集，以重新训练DAIN进近器的网络，以满足冠状动脉造影的应用场景。同时，我们还重新训练了其他几种深度学习算法，以比较和分析这些帧插值算法的结果。证明了使用视频帧插值算法合成连续和清晰的高帧率冠状动脉造影视频的可行性。我们证明了通过内插帧合成的高帧率冠状动脉血管造影视频可以帮助医生更准确地确定冠状动脉的血管状态。未来我们计划提高帧插值算法的工作效率，使其能够满足冠状动脉造影术中的实时要求。借助该技术，医生可以显着降低冠状动脉造影剂的浓度和剂量。血管造影并获得出色的诊断结果，从而减少或避免了高浓度大剂量造影剂注射的副作用，尤其是对于肾功能下降的患者

%\subsection{\label{sec:citeref} References}
%\ifCLASSOPTIONcaptionson
%\newpage
%\fi
%\subsection{\label{sec:citeref} References}

% \subsubsection{Citations}

% \paragraph{Syntax}
 
 \begin{quotation}\flushleft\leftskip1em

 \end{quotation}\noindent

 \begin{quotation}\flushleft\leftskip1em
 	
\end{quotation}\noindent

%Dear Editors:
%We would like to submit the enclosed manuscript entitled “Reducing the contrast agent dosage via deep learning-based video interpolation in cardio-angiography”, which we wish to be considered for publication in “IEEE Transactions on Biomedical Engineering”. 
%In this work, we innovatively apply deep learning-based video frame interpolation algorithms to coronary angiography videos.  We make a new coronary angiography dataset for video interpolation algorithm  and prove that using the retrained DAIN model can synthesize continuous and clear high frame rate coronary angiography video.With the help of this technology, doctors can significantly reduce the concentration and dose of contrast agents during coronary angiography and obtain good diagnostic results, thereby reducing or avoiding the side effects of high-concentration large-dose contrast agent injections, especially for patients with reduced kidney function.
%I would like to declare on behalf of my co-authors that this manuscript is entirely original,  has not been copyrighted,  published,  submitted,  or accepted for publication elsewhere. 
%We deeply appreciate your consideration of our manuscript, and we look forward to receiving comments from the reviewers. If you have any queries, please don’t hesitate to contact me at the address below.
%Thank you and best regards.
%Yours sincerely,
%Xiaolei Yin
%Corresponding author:
%Name: Dongxue Liang
%E-mail: liang_laurel@tsinghua.edu.cn

%\paragraph{Eliding }

%\paragraph{The f}

\end{document}